\documentstyle[amssymb,aps,floats,twocolumn,psfig,epsf]{revtex}
\begin{document}
\title{Electronic self-energy and triplet pairing fluctuations in the vicinity of a
ferromagnetic instability in 2D systems: the quasistatic approach}
\author{A. A. Katanin$^{a,b}$}
\address{$^a$Max-Planck-Institut f\"ur Festk\"orperforschung, 70569, Stuttgart,
Germany\\
$^b$Institute of Metal Physics, 620219 Ekaterinburg, Russia}
\maketitle

\begin{abstract}
{The self-energy, spectral functions and susceptibilities of 2D systems with
strong ferromagnetic fluctuations are considered within the quasistatic
approach. T}he self-energy at low temperatures $T$ has a non-Fermi liquid
form in the energy window $|\omega |\lesssim \Delta _0$ near the Fermi
level, where $\Delta _0$ is the ground-state spin splitting for magnetically
ordered ground state, and $\Delta _0\propto T^{1/2}\ln ^{1/2}(v_F/T)$ in the
quantum critical regime ($v_F$ is the Fermi velocity). Spectral functions
have a two-peak structure at finite $T$ above the magnetically ordered
ground state, which implies quasi-splitting of the Fermi surface in the
paramagnetic phase in the presence of strong ferromagnetic fluctuations. The
triplet pairing amplitude in the quasistatic approximation increases with
increasing correlation length; at low temperatures $T\ll \Delta _0$ the
vertex corrections become important and the Eliashberg approach is not
justified. The results for the spectral properties and susceptibilities in
the quantum critical regime near charge- (spin-) instabilities with large
enough correlation length $\xi \gg (T/v_F)^{-1/3}$ are obtained.
\end{abstract}

\section{Introduction}

Anomalous non-Fermi-liquid behavior of correlated low-dimensional electron
systems has attracted much attention during the last decade. This behavior
is usually connected with the violation of the quasiparticle (qp) concept in
some energy window around the Fermi level. A prominent example is the
pseudogap phenomenon observed in underdoped high-T$_c$ compounds \cite{ARPES}%
. While antiferromagnetic (AFM) fluctuations may be responsible for
non-Fermi-liquid behavior and superconducting pairing in cuprates, there is
number of systems where ferromagnetic (FM) fluctuations may play an
important role. In particular, FM fluctuations may be important in some
triplet superconductors, such as UGe$_2,$ ZrZn$_2$ and Sr$_2$RuO$_4$. These
systems motivate studies of electronic properties in the vicinity of a FM
instability and their influence on the triplet superconductivity.

Although many results exist for the electronic properties in the vicinity of
an AFM state \cite{Pines,Schmalian,SF,FLEX,TPSC,DCA,KK}, much less is known
about the evolution of quasiparticle properties near the FM instability. An $%
\varepsilon ^{2/3}$ energy dependence of the self-energy at the quantum
critical point (QCP) can be derived from calculations in the context of
gauge field theories \cite{Lee}, the phase separation problem \cite
{Castellani}, and the Pomeranchuk instability \cite{MetznerSE}, which are
expected to have the same structure of self-energy corrections as a FM
instability. The breakdown of the qp concept at the QCP is even more
apparent at finite temperatures. It was demonstrated for fermions
interacting with a gauge field that the imaginary part of the self-energy in
a non-self-consistent calculation diverges at the Fermi level at $T>0$ as a
consequence of the divergence of the gauge field propagator at zero momentum
and frequency \cite{Lee}. Similar behavior induced by the divergence of the
static uniform spin susceptibility $\chi ({\bf 0},0)$ can be expected for
the zero-momentum particle-hole instabilities of fermion systems with
short-range interactions. This behavior can be especially pronounced in the
renormalized classical (RC) regime \cite{Sachdev}, where the correlation
length $\xi $ is exponentially large.

The self-energy and the spectral functions in the RC regime in the vicinity
of a FM instability were previously studied within the two-particle
self-consistent (TPSC) approximation \cite{TPSC1,KKI}, one-loop functional
renormalization group (fRG), and Ward-identity approaches \cite{KKI}. It was
argued, that spectral functions have two-peak structure analogously to the
vicinity of an AFM instability\cite{TPSC}. Contrary to the situation in the
vicinity of an AFM instability, however, the abovementioned two-peak
structure of the spectral functions does not imply strong suppression of the
density of states at the Fermi level, but leads to the quasi-splitting of
the Fermi surface at low $T$ already in the paramagnetic phase \cite{KKI}.
While the treatment based on the TPSC\ and one-loop fRG approaches does not
account for the feedback of the self-energy effects, the analysis of the
self-energy and vertex corrections using Ward identities has shown \cite{KKI}
that these two types of effects almost cancel each other, and therefore
resulting spectral functions closely resemble their form in
non-self-consistent approaches.

These anomalous spectral properties may have a profound effect on the
triplet superconductivity. One can expect that due to strong FM fluctuations
in the RC regime the triplet pairing will be mostly enhanced at the new
preformed Fermi surfaces. Anomalous spectral properties may have important
influence also in quantum-critical (QC) regime \cite{Sachdev}, where the
quasi-splitting of the FS is absent. Previous investigations of this regime 
\cite{Bedell,CC,Gorkov} neglected vertex corrections, which may be important
for large enough correlation length.

To consider anomalies of electronic properties and their impact on the
triplet superconductivity, one needs a tool which is able to consider
self-energy and vertex corrections on the same foot. The abovementioned
self-consistent treatment of self-energy and vertex corrections near a FM
instability was performed only to first order in $1/M$, $M$ being the number
of spin components ($M=3$ for the Hubbard model). It appears important to
investigate spectral properties in the vicinity of FM instability beyond the
leading order in $1/M$ to verify whether the near-cancellation of the
self-energy and vertex corrections persists also in higher orders of the $%
1/M $ expansion and to investigate the effect of the anomalous properties on
the triplet superconductivity.

Due to an almost static character of spin fluctuations at large correlation
length, a useful nonperturbative tool for calculation of the spectral
properties and susceptibilities in this case is the quasistatic approach.
This approach was originally proposed for the summation of diagrammatic
series for the self-energy of one-dimensional (1D) systems in the vicinity
of the charge-density wave instability \cite{Sadovskii} and further
developed for 2D systems in the vicinity of an AFM instability \cite
{Schmalian,Sadovskii1}. The quasistatic approach allows to sum up the most
important series of static contributions to the self-energy and interaction
vertices. This approach becomes exact in the limit $\xi \rightarrow \infty
,\ $and can be applied to study spectral properties in the RC and QC regime,
provided that the correlation length is sufficiently large, $(\xi
/a)^{-1}\ll (T/v_F)^{1/3},$ where $v_F$ is the Fermi velocity, $a$ is the
lattice constant (the latter criterion follows from the comparison of static 
contributions to the scattering rate $\sim T\xi /a$ with the dynamic 
contributions proportional to $v_F(T/v_F)^{2/3},$ cf. Ref. \cite{KKI}). 
Although the quasistatic approach was applied
previously to systems in the vicinity of a FM instability \cite{Month}, only
the form of spectral functions was analyzed, the self-energy, magnetic and
triplet pairing susceptibilities being not investigated.

In the present paper we apply the quasistatic approach to 2D systems with
nonsingular density of states, which are on the verge of a ferromagnetic
instability, to study spectral properties and the possibility of triplet
pairing in these systems. In Sect. II we concentrate on the analytical
results for spectral properties and susceptibilities for linear electronic
dispersion at $\xi \rightarrow \infty $ and compare these results with the
results at finite correlation length. In Sect. III we consider the
two-particle properties: magnetic susceptibility and the susceptibility with
respect to triplet pairing. In Sect. IV we discuss the application of the
results to the quantum-critical regime. Finally, in Sect. V we summarize
main results of the paper.

\section{Spectral properties in the vicinity of ferromagnetic instability}

We consider a spin-fermion model \cite{Pines,Schmalian} 
\begin{eqnarray}
Z[\eta ] &=&\int D[c,c^{\dagger }]D[{\bf S}]\exp (-{\cal S}[c,{\bf S},\eta ])
\nonumber \\
{\cal S}[c,{\bf S},\eta ] &=&\sum_k\left[ c_{k\sigma }^{\dagger }(i\omega
_n-\varepsilon _{{\bf k}})c_{k\sigma }-(c_{k\sigma }^{\dagger }\eta
_{k\sigma }+\eta _{k\sigma }^{\dagger }c_{k\sigma })\right]  \nonumber \\
&&\ +T^{-1}\sum_q(\chi _q^{-1}+U^2\Pi _q)({\bf S}_q{\bf S}_{-q})  \nonumber
\\
&&\ +UT^{-1}\sum_{kk^{\prime }}{\bf S}_{k-k^{\prime }}\sigma _{\sigma \sigma
^{\prime }}c_{k\sigma }^{\dagger }c_{k^{\prime }\sigma ^{\prime }}
\label{Hf}
\end{eqnarray}
where $q=({\bf q},i\omega _n)$ and similar for $k,$ $\omega _n=(2n+1)\pi T$
are fermionic Matsubara frequencies, $\varepsilon _{{\bf k}}$ is the
electronic dispersion, $\sigma _{\sigma \sigma ^{\prime }}$ are Pauli
matrices, $\chi _q=\Pi _q/(1-U\Pi _q)$ is the dynamical spin susceptibility, 
\begin{equation}
\Pi _{{\bf q},i\omega _n}=\sum_{{\bf k}}\frac{f(\varepsilon _{{\bf k}+{\bf q}%
})-f(\varepsilon _{{\bf k}})}{i\omega _n-\varepsilon _{{\bf k}+{\bf q}%
}+\varepsilon _{{\bf k}}}
\end{equation}
is the bare polarization operator, $f(\varepsilon )$ is the Fermi
distribution function, $U$ is the strength of the interaction of electrons
with the collective magnetic excitations, the lattice constant $a=1$. 
Although this model was originally
proposed as phenomenological model for systems with strong AFM fluctuations,
it can be applied for systems with strong FM fluctuations as well. Generally
speaking, the interaction $U$ differs from the bare on-site Coulomb
repulsion because of contributions of the channels of electron-electron
scattering different from particle-hole one, therefore $U$ should be
considered as an effective interaction. The counterterm proportional to $%
U^2\Pi _q$ keeps the renormalized spin-spin propagator equal to $\chi _q$
(see below)$.$ The rigorous derivation of the model (\ref{Hf}) from the
microscopic Hubbard model will be considered elsewhere \cite{OurP}.

Integrating out fermions from (\ref{Hf}), we obtain 
\begin{eqnarray}
Z[\eta ] &=&\int D[{\bf S}]\exp (-{\cal S}_{eff}[{\bf S},\eta ])  \label{act}
\\
{\cal S}_{eff}[{\bf S},\eta ] &=&T^{-1}\sum_q\chi _q^{-1}{\bf S}_q{\bf S}%
_{-q}-\ln \det [-G_{kk^{\prime }}^{-1}({\bf S})]  \nonumber \\
&&+U^2T^{-1}\sum_q\Pi _q{\bf S}_q{\bf S}_{-q}+\sum_{kk^{\prime }}\eta
_k^{\dagger }G_{kk^{\prime }}^{-1}({\bf S})\eta _{k^{\prime }}  \nonumber
\end{eqnarray}
where 
\begin{equation}
G_{kk^{\prime }}^{-1}({\bf S})=(i\omega _n-\varepsilon _{{\bf k}})\delta
_{\sigma \sigma ^{\prime }}\delta _{kk^{\prime }}+U{\bf S}_{k-k^{\prime
}}\sigma _{\sigma \sigma ^{\prime }}
\end{equation}
In the following we expand $\ln \det (-G^{-1})$ in Eq. (\ref{act}) in powers
of ${\bf S}$ and retain only quadratic term, which is exactly cancelled by
the counterterm introduced in Eq. (\ref{Hf}) (so that the propagator of the
field ${\bf S}$ remains equal to $\chi _q$), the relevance of higher-order
terms is discussed below.

We start investigation of the functional (\ref{act}) with the consideration
of the limit $\xi \rightarrow \infty ,$ where spin fluctuations are
especially strong. At $\xi \rightarrow \infty $ the susceptibility $\chi
_q=\chi ({\bf q},i\omega _n)$ is divergent at ${\bf q}=0,$ $\omega _n=0.$
Since the momentum transfer for the scattering on these most singular
magnetic fluctuations is small, it can be neglected in the electronic Geen
functions. Sums over internal momenta in all diagrams are applied then only
to the propagators $\chi _q$ of the spin field ${\bf S}_q,$ so that the
action (\ref{act}) in $\xi \rightarrow \infty $ limit can be reduced to an
effective action which contains only one fluctuating field ${\bf S\equiv S}%
_{q=0}$ (cf. Refs. \cite{Sadovskii,Schmalian,Sadovskii1}) 
\begin{eqnarray}
&&\ Z[\eta ]
\begin{array}{c}
=
\end{array}
\int d^3{\bf S}\exp (-{\cal S}_{eff}[{\bf S},\eta ])  \label{actm} \\
&&\ {\cal S}_{eff}[{\bf S},\eta ]
\begin{array}{c}
=
\end{array}
\frac{3U^2}{2\Delta _0^2}S^2-T\sum_{{\bf k},i\omega _n}\frac 1{(i\omega
_n-\varepsilon _{{\bf k}})^2-U^2S^2}  \nonumber \\
&&\ \ \ \ \ \ \ \ 
\begin{array}{c}
\times 
\end{array}
\eta _{{\bf k},i\omega _n}^{\dagger }\left( 
\begin{array}{cc}
i\omega _n-\varepsilon _{{\bf k}}+US^z & US^{-} \\ 
US^{+} & i\omega _n-\varepsilon _{{\bf k}}-US^z
\end{array}
\right) \eta _{{\bf k},i\omega _n}  \nonumber
\end{eqnarray}
where $S^{\pm }=S^x\pm iS^y$. The effective propagator of the field ${\bf S}$%
, $\Delta _0^2/(3U^2)$, is determined by the average (local) spin
susceptibility 
\begin{equation}
\Delta _0^2=\frac{3U^2T}2%
\mathop{\displaystyle \sum }
\limits_{{\bf q}}\chi ({\bf q},0)  \label{dd}
\end{equation}
For an ordered ground state $\Delta _0^2$ is almost independent of
temperature at low $T$ and its $T\rightarrow 0$ limit is equal to the square
of the ground-state spin splitting. At the same time, in the QC regime we
have $\Delta _0^2\propto T\ln (v_F/T)$, in this case the effect of finite
correlation length should be also accounted for. Due to neglection of terms
which are of higher order $n>2$ in spin operators and proportional to $\rho
_0^{(n-2)}(\varepsilon ),$ [$\rho _0(\varepsilon )$ being noninteracting
density of states], the generating functional (\ref{actm}) is valid only for
regular $\rho _0(\varepsilon )$, which is smooth enough in the vicinity of
the Fermi level to satisfy $U^n\rho _0^{(n)}(\varepsilon )\ll \rho
_0(\varepsilon )$ at $n>0$ \cite{OurP}. 

Similar to Ref. \cite{Schmalian} we generalize the action (%
\ref{actm}) to $M$-component field ${\bf S}=(S_1...S_M),$ $M=3$ for the
model (\ref{Hf}); this generalization also allows one to consider a charge
instability with $M=1$. The results for the observable quantities are found
by differentiation of partition function over the source fields $\eta $ and
are expressed as integrals over the field ${\bf S}$ of some functions $f(S)$.

For the electronic Green function at $\xi \rightarrow \infty $ we obtain the
result 
\begin{eqnarray}
G(\overline{\omega }) &=&\frac{\delta ^2Z}{\delta \eta ^{\dagger }\delta
\eta }=\int d^M{\bf S}\frac{\overline{\omega }}{\overline{\omega }^2-U^2S^2}%
\exp (-{\cal S}_{eff}[{\bf S},0])  \nonumber \\
&=&\frac 1{\overline{\omega }}\left( -\frac{M\,\overline{\omega }^2}{2\Delta
_0^2}\right) ^{M/2}e^{-\frac{M\,\overline{\omega }^2}{2\Delta _0^2}}\Gamma
\left( 1-\frac M2,-\frac{M\overline{\omega }^2}{2\Delta _0^2}\right) 
\label{Gwr}
\end{eqnarray}
which depends on $\overline{\omega }=\omega -\varepsilon _{{\bf k}}$ only, $%
\Gamma \left( a,x\right) $ is the incomplete Gamma function. The result (\ref
{Gwr}) is similar to previous result in the vicinity of an AFM instability 
\cite{Schmalian}. The electronic self-energy is given by 
\begin{equation}
\Sigma (\overline{\omega })=\overline{\omega }-G^{-1}(\overline{\omega })
\label{Swr}
\end{equation}
The retarted Green function and self-energy on the real axis are obtained by
the replacement $\omega \rightarrow \omega +i0^{+}.$ For the following
analysis it is convenient to introduce a one-particle irreducible (1PI)
vertex 
\begin{eqnarray}
\gamma ^a({\bf k},i\omega ) &=&G_k^{-2}\sum_{{\bf k}^{\prime }\sigma
^{\prime \prime }\sigma ^{\prime \prime \prime }}\int_0^\beta d\tau
\int_0^\beta d\tau ^{\prime }e^{-i\omega (\tau -\tau ^{\prime })}\sigma
_{\sigma ^{\prime \prime }\sigma ^{\prime \prime \prime }}^a  \label{gamma}
\\
&&\times \lim_{{\bf q}\rightarrow 0}\chi _{{\bf q}}^{-1}\langle T[c_{{\bf k}%
^{\prime }\sigma ^{\prime \prime }}^{\dagger }(\tau )c_{{\bf k}^{\prime }+%
{\bf q,}\sigma ^{\prime \prime \prime }}(\tau ^{\prime })S_{{\bf q}%
}^a(0)]\rangle _{\text{1PI}}  \nonumber
\end{eqnarray}
where $a=x,y,z$. Similar to the Green function (\ref{Gwr}), $\gamma ({\bf k}%
,\omega )$ depends on $\overline{\omega }$ only: $\gamma ^a({\bf k},\omega )=
$ $\gamma (\overline{\omega }).$ The function $\gamma (\overline{\omega })$
can be obtained from the exact Dyson relation, connecting the vertex and the
self-energy (see, e.g., Ref. \cite{HertzEdw}), which at $\xi \rightarrow
\infty $ takes rather simple form 
\begin{equation}
\Sigma (\overline{\omega })=\frac{\Delta _0^2\gamma (\overline{\omega })}{%
\overline{\omega }-\Sigma (\overline{\omega })}  \label{Dy}
\end{equation}
The quantities $G(\omega )$, $\Sigma (\omega )$, and $\gamma (\omega )$
determined by Eqs. (\ref{Gwr})-(\ref{Dy}) can be considered as perturbation
series in $\Delta _0\propto U$. The corresponding lowest-order coefficients
obtained from Eqs. (\ref{Gwr}) and (\ref{Swr}) are 
\begin{eqnarray}
G(\omega ) &=&\frac 1\omega +\frac{\Delta _0^2}{\omega ^3}+\frac{\left(
2+M\right) \,\Delta _0^4}{M\,\omega ^5}+{\cal O}(\Delta _0^5)  \nonumber \\
\Sigma (\omega ) &=&\frac{\Delta _0^2}\omega +\frac{2\Delta _0^4}{M\,\omega
^3}+\frac{2\,\left( 4+M\right) \,\Delta _0^6}{M^2\,\omega ^5}+{\cal O}%
(\Delta _0^7)  \nonumber \\
\gamma (\omega ) &=&1+\frac{\left( 2-M\right) \Delta _0^2}{M\omega ^2}+\frac{%
2\left( 4-M\right) \Delta _0^4}{M^2\omega ^4}+{\cal O}(\Delta _0^5)
\label{pert}
\end{eqnarray}
The coefficients of the series in $\Delta _0$ can be found also directly
from a diagram technique (we have verified correspondence of several
lowest-order terms).

The perturbation series (\ref{pert}) breaks down at frequencies $|\omega
|\lesssim \Delta _0$, although nonperturbative results (\ref{Gwr})-(\ref{Dy}%
) can be used to analyze physical properties in this frequency range as
well. In particular, for $M>2$ we find 
\begin{equation}
\text{Re}\Sigma (\omega )\simeq 
\begin{array}{cc}
(M-2)\Delta _0^2/M\omega , & |\omega |\ll \Delta _0.
\end{array}
\end{equation}
Comparing this result with the perturbation theory result (\ref{pert}) one
observes, that Re$\Sigma (\omega )$ has a nontrivial crossover with the
reduction of the number of spin components at low frequencies. Such a
crossover is similar to that for the spin-spin correlation function in 2D
and quasi-2D generalized Heisenberg model with $O(M)/O(M-1)$ symmetry \cite
{CSY,Quasi2D}. In the crossover region $|\omega |\sim \Delta _0$ the real
part of the self-energy $\Sigma (\omega )$ is only weakly $\omega $%
-dependent. The imaginary part of the self-energy at small $|\omega |$ and $%
M>2$ reads 
\begin{eqnarray}
\text{Im}\Sigma (\omega ) &\simeq &-A_M(2/M)^{1/2}(M/2-1)^2\Delta _0
\label{M12} \\
&&\times (M\omega ^2/2\Delta _0^2)^{(M-3)/2}  \nonumber \\
&&-\pi (M-2)\Delta _0^2\delta (\omega )/M,\,\,\,\,\,\,
\begin{array}{c}
|\omega |\ll \Delta _0
\end{array}
,  \nonumber
\end{eqnarray}
where $A_M=\pi /(M/2-1)!$ for even $M$ and $A_M=\Gamma (1-M/2)$ for odd $M$.
For the charge instability case ($M=1$) and small $|\omega |\ll \Delta _0$
we obtain

\begin{eqnarray}
\text{Re}\Sigma (\omega )\simeq 2\omega /\pi ^2;\text{ Im}\Sigma (\omega
)\simeq -\sqrt{2/\pi }\Delta _0\,\,\,(M=1)  \nonumber
\end{eqnarray}
The overall frequency dependence of $\Sigma (\omega )$, $\gamma (\omega ),$
and the spectral function $A(\omega )=-$Im$G(\omega )/\pi $ at $\xi
\rightarrow \infty $ calculated using Eqs. (\ref{Gwr})-(\ref{Dy}) for
different $M$ is shown and compared with the results of $1/M$ expansion of
Ref. \cite{KKI} in Fig. 1. One can see that at $M=3$ (the same behavior
takes place for all $M>1$) the real part of the self-energy has (infinite)
positive slope at the Fermi level, where the imaginary part of self-energy
has $\delta $-like singularity, and the spectral function has two-peak
structure$.$ These features are in qualitative agreement with previous
results of the first-order $1/M$ analysis \cite{KKI}. They arise as a result
of strong FM fluctuations and violate the quasiparticle concept near the
Fermi level. It was argued in Ref. \cite{KKI} that the two-peak structure of
the spectral function, together with its dependence on $\omega -\varepsilon
_{{\bf k}}$ implies pre-formation of the two new Fermi surfaces already in
the paramagnetic phase (so called quasi-splitting of the FS), the same
arguments can be applied to the results of quasistatic approach.

The main difference of the results of quasistatic approach from the results
of $1/M$ expansion \cite{KKI} is in partial transfer of the spectral weight
from the peaks of the two peak structure to small $\omega $-region, where
the spectral weight in the result (\ref{Gwr}) is small, but finite. From Eq.
(\ref{Gwr}) we find $A(\omega )\sim |\omega |^{M-1}$ at small $\omega .$ The
nonanalytical dependence of $A(\omega )$ on $M$ explains why this behavior
is not captured by $1/M$ expansion.

\begin{figure}[t!]
\psfig{file=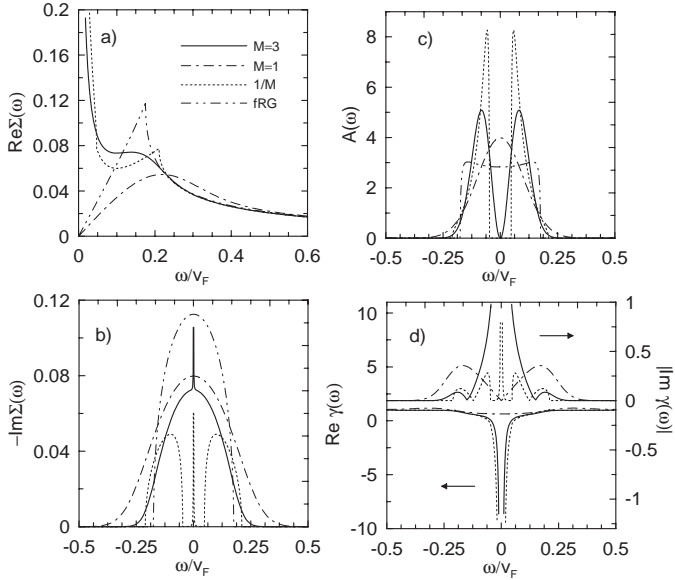,width=90mm,silent=} \vspace{2mm}
\caption{The real and imaginary parts of the self-energy (a,b), the spectral
function (c), and the vertex function $\gamma $ (d) in the quasistatic
approximation at $\xi \rightarrow \infty ,$ $M=1$ (dot-dashed lines) and $%
M=3 $ (solid lines) as a function of $\overline{\omega }=\omega -\varepsilon
_{{\bf k}}$ for $\Delta _0/v_F=0.1,$ compared to the result of $1/M$
expansion (dotted lines) and the one-loop functional RG approach for
spin-fermion model (dot-dot dashed line) for $M=3$.}
\label{fig:Fig1}
\end{figure}

For $M=1$ (charge instability case) the imaginary part of the self-energy is
finite at the Fermi level [see Eq. (\ref{M12})] and the spectral function
has one-peak structure. It does not imply, however, the validity of the
quasiparticle concept, since the real part of the self-energy has positive
slope at the Fermi level, which invalidates quasiparticle picture. Note that
vertex corrections are finite in this case, and, therefore, are not as
important, as for $M>1.$ Since the long-range order exists for $M=1$ also at
finite $T,$ these results are applicable only in a narrow critical regime
near the transition temperature, but have also some implication for
quantum-critical region, as discussed in Sect. IV. The behavior of $G(\omega
)$, $\Sigma (\omega )$ and $\gamma (\omega )$ at $M=2$ (XY-type symmetry) is
very similar to that for $M=3$, except for additional logarithmic
corrections.

It is instructive to compare the results (\ref{Gwr})-(\ref{Dy}) for the
self-energy and vertex at $\xi \rightarrow \infty $ with the corresponding
results of recently proposed functional renormalization-group approach for
the boson-fermion model\cite{Schutz}. Since the momenta integrations and
frequency summations in Feynman diagrams are restricted at large $\xi $ to i$%
\omega _n=0$ and the near vicinity of $q=0$, neither frequency, nor momentum
cutoff of electronic or bosonic degrees of freedom are convenient for this
problem. Instead, we impose the temperature cutoff on the electronic Green
function (the correlation length is kept fixed, so that the bosonic
propagator does not acquire temperature dependence). We also combine this
scheme with the one-particle self-consistent modification of fRG equations 
\cite{MyWard}, which allows for a correct treatment of self-energy effects
to one-loop order. The resulting one-loop fRG equations at $\xi \rightarrow
\infty $ read 
\begin{eqnarray}
\frac{d\Sigma (\omega )}{d\omega } &=&\Delta _0^2\gamma ^2(\omega )\frac{%
dG(\omega )}{d\omega }  \label{GR} \\
\frac{d\gamma (\omega )}{d\omega } &=&-2\frac{M-2}M\Delta _0^2\gamma
^3(\omega )G(\omega )\frac{dG(\omega )}{d\omega }  \nonumber
\end{eqnarray}
where $G(\omega )=[\omega -\Sigma (\omega )]^{-1}.$ We compare the solution
of Eqs. (\ref{GR}) with the result of quasistatic approach (\ref{Gwr}) in
Fig. 1. One can see that the one-loop fRG equations describe very accurately
the perturbative regime $|\omega |\gtrsim \Delta _0,$ but their description
breaks down in the strong-coupling regime $|\omega |\lesssim \Delta _0.$

To study the effect of finite correlation length, we employ an ansatz for
the nonuniform magnetic susceptibility 
\begin{equation}
\chi ({\bf q},0)=\frac A{q^2+\xi ^{-2}}  \label{hi11}
\end{equation}
Note that this ansatz neglects recently found nonanalytic corrections \cite
{Nonan} and is, therefore, valid above the characteristic temperature $%
T_X\sim U^2/v_F,$ where these corrections become important. At finite $\xi $
the quasistatic approach can be applied when static contributions to the
self-energy and vertices are dominating near the Fermi level, i.e. at $\xi
^{-1}\ll (T/v_F)^{1/3},$ as discussed in the introduction. This condition is
satisfied, in particular, in the RC regime.

The generalization of the quasistatic approach to the susceptibility ansatz (%
\ref{hi11}) is considered in Appendix. The result (\ref{Gwr}) is to be
replaced at finite $\xi $ by an integral recursion relations for the
electronic self-energy $\Sigma (\omega )=\Sigma _1(\omega )$ and vertex $%
\gamma (\omega )=\gamma _1(\omega )$, 
\begin{eqnarray}
\Sigma _j(\omega ) &=&\frac{\Delta _0^2c_j}{2\ln \xi }\int_{-\infty }^\infty 
\frac{da}{\sqrt{a^2+\xi ^{-2}}}G_{j+1}(\omega -v_Fa)  \label{gs1} \\
\gamma _j(\omega ) &=&1-\frac{\Delta _0^2r_j}{2\ln \xi }\int_{-\infty
}^\infty \frac{da}{\sqrt{a^2+\xi ^{-2}}}  \nonumber \\
&&\ \ \ \ \ \times \gamma _{j+1}(\omega -v_Fa)G_{j+1}^2(\omega -v_Fa)
\label{gj1}
\end{eqnarray}
where $G_j(\omega )=[\omega -\Sigma _j(\omega )]^{-1}$, $c_j=j/M$ (even $j$%
), $c_j=(j+M-1)/M$ (odd $j$); $r_j=j/(M-2)$ (even $j$), $r_j=(M-2)(j+M-1)/M^2
$ (odd $j$), $G_\infty (\omega )=1/\omega ,$ and $\gamma _\infty (\omega )=1.
$ The most important contributions to integrals in Eqs. (\ref{gs1}) and (\ref
{gj1}) at $\xi \rightarrow \infty $ come from a narrow vicinity of the point 
$a=0,$ and these equations reduce to the continuous fraction representation
of the gamma-function in Eq. (\ref{Gwr}). At the same time, at finite $\xi $
the Eqs. (\ref{gs1}) and (\ref{gj1}) have to be solved numerically.

The results for the self-energy and the vertex are shown for different
values of $\xi $ and compared with the results for $\xi \rightarrow \infty $
in Fig. 2. In agreement with previous analysis \cite{KKI}, the real part of
the self-energy acquires a large positive slope at the Fermi level, $%
\partial $Re$\Sigma /\partial \varepsilon \sim T\xi ^2.$ The $\delta $%
-function singularity at $\xi \rightarrow \infty $ in the imaginary part of
the self-energy is replaced by the lorentian-like form of the imaginary part
with $|$Im$\Sigma (0)|\sim T\xi ,$ so that the quasiparticle picture is
invalid at finite $\xi $ as well. With decreasing correlation length the
structure of the spectral function changes from the two- to one-peak form at 
$\xi ^{-1}\sim \Delta _0/v_F.$ Contrary to $\xi \rightarrow \infty $ limit,
the vertex remains finite at finite $\xi .$

At $M=1$ the imaginary part of the self-energy, which was finite at $\xi
\rightarrow \infty ,$ is determined by $|$Im$\Sigma (0)|\sim \min (\Delta
_0,T\xi )$. The qp picture is violated in this case as well, since the slope
of Re$\Sigma $ is positive, $\partial $Re$\Sigma /\partial \varepsilon \sim
1.$

\begin{figure}[t!]
\psfig{file=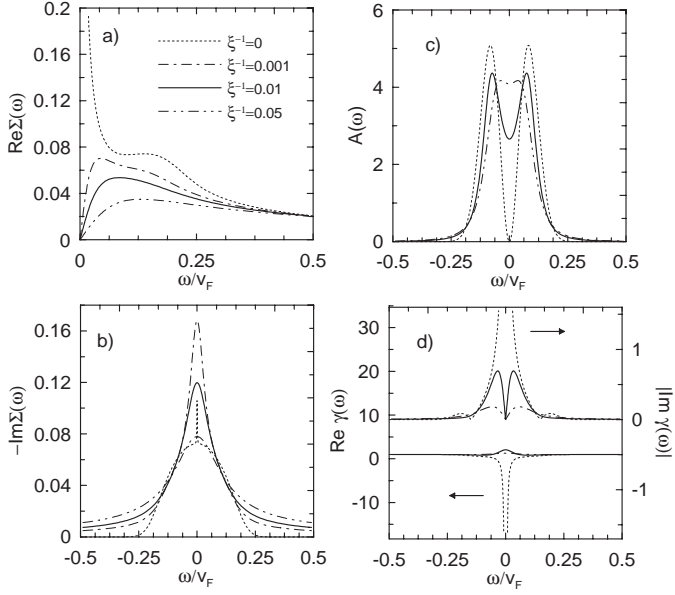,width=90mm,silent=} \vspace{2mm}
\caption{The real and imaginary parts of the self-energy (a,b), the spectral
function (c), and the vertex function $\gamma $ (d) in the quasistatic
approximation at $M=3$, $\Delta _0/v_F=0.1$, and different values of the
correlation length.}
\label{fig:Fig2}
\end{figure}

\section{Two-particle properties}

Now we discuss two-particle properties. First we consider static uniform
spin susceptibility $\phi _{\text{ph}}.$ According to Ref. \cite{HertzEdw},
this susceptibility can be expressed through the irreducible in the
particle-hole channel susceptibility $\phi _{\text{ph,0}}$ via the relation 
\begin{equation}
\phi _{\text{ph}}=\phi _{\text{ph,0}}/(1-U\phi _{\text{ph,0}})
\end{equation}
which is similar to the random phase approximation result with the
difference that $\phi _{\text{ph,0}}$ includes the self-energy and vertex
corrections. Using the definition of the irreducible vertex, Eq. (\ref{gamma}%
), we find 
\begin{eqnarray}
\phi _{\text{ph,0}} &=&\int d\varepsilon \rho _0(\varepsilon )\Phi
(\varepsilon )  \nonumber \\
\ \Phi (\varepsilon ) &=&-T\sum_{i\omega _n}\phi (i\omega _n-\varepsilon
),\,\,\phi (z)=G^2(z)\gamma (z)  \label{fi}
\end{eqnarray}
The necessary condition for the existence of a ferromagnetic instability is $%
\phi _{\text{ph},0}>0;$ the function $\Phi (\varepsilon )$ characterizes the
relative weight of states with different energy in $\phi _{\text{ph,0}}$.
The plot of the function $\phi (z)$ in the complex plane at $\xi \rightarrow
\infty $ is shown in Fig. 3a (the plot of this function at finite $\xi $
looks similarly). At $\xi \rightarrow \infty $ the contribution of regions $%
|\omega |<\Delta _0$ and $|\omega |>\Delta _0$ to $\Phi (\varepsilon )$ have
different signs and compensate each other for all $\varepsilon $, except for 
$|\varepsilon |<\Delta _0,$ where $\Phi (\varepsilon )$ is maximum (Fig.
3b). Therefore, irreducible susceptibility depends on the details of the
density of states only in the energy range $|\varepsilon |<\Delta _0$. One
can see that for regular densities of states, which are not strongly
suppressed in this energy range, the condition $\phi _{\text{ph,}0}>0$ can
be easily fulfilled. Stronger criterium of stability of ferromagnetism $%
\partial ^2\phi _q/\partial q^2<0$ is studied in detail in a forthcoming
paper \cite{OurP}.

To investigate the static magnetic susceptibility at finite $\xi ,$ we
suppose that the temperature dependence of the correlation length is given
by $\xi =\exp (T^{*}/T)$ where $T^{*}$ is the crossover temperature to the
renormalized classical regime$.$ The function $\Phi (\varepsilon )$ for
different values of $\xi $ is shown in Fig. 3b. At not too large correlation
length the energy range which contributes to $\phi _{\text{ph,0}}$ is spread
to $|\varepsilon |>\Delta _0$ as well. At the same time, the total area
under $\Phi (\varepsilon )$ changes rather weakly and, therefore, one can
expect weak dependence of the irreducible susceptibility on the correlation
length $\xi .$

\begin{figure}[t!]
\psfig{file=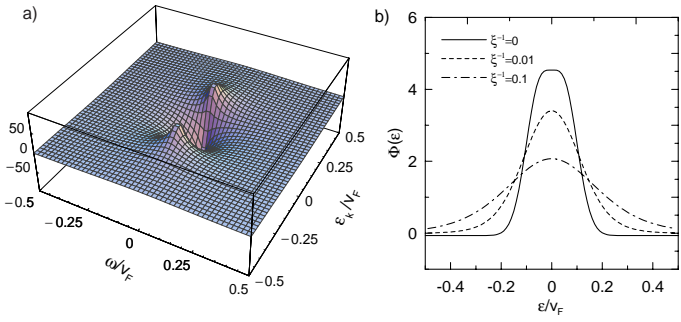,width=90mm,silent=} \vspace{2mm}
\caption{(a) The plot of the function $\phi (i\omega -\varepsilon )$ at $\xi
\rightarrow \infty $ and $\Delta _0=0.1 v_F$. (b) The function $\Phi
(\varepsilon )$ (which determines the uniform static irreducible spin
susceptibility according to Eq. (\ref{fi})) at $\Delta _0=0.1 v_F$, $%
T^{*}=0.25 v_F$, and different values of the correlation length.}
\label{fig:Fig3}
\end{figure}

The susceptibility with respect to triplet pairing 
\begin{eqnarray}
\phi _{\text{pp},\text{tr}}^a &=&\sum_{{\bf kk}^{\prime }}\sum_{\sigma
\sigma ^{\prime }\sigma ^{\prime \prime }\sigma ^{\prime \prime \prime
}}A_{\sigma \sigma ^{\prime }}^aA_{\sigma ^{\prime }\sigma ^{\prime \prime
}}^a\int_0^\beta d\tau   \label{hipair} \\
&&\ \ \ \ \ \times \langle T[c_{{\bf k,}\sigma }^{\dagger }(\tau )c_{-{\bf k,%
}\sigma ^{\prime }}^{\dagger }(\tau )c_{{\bf k}^{\prime }{\bf ,}\sigma
^{\prime \prime }}(0)c_{-{\bf k}^{\prime }{\bf ,}\sigma ^{\prime \prime
\prime }}(0)]\rangle   \nonumber
\end{eqnarray}
where $A_{\sigma \sigma ^{\prime }}^a=(\sigma ^a\sigma ^y)_{\sigma \sigma
^{\prime }}$ can be considered in a similar way. It is convenient to
represent it in the form 
\begin{eqnarray}
\phi _{\text{pp},\text{tr}}^a &=&T\sum_{{\bf k,}i\omega _n}\sum_{\sigma
\sigma ^{\prime }}A_{\sigma \sigma ^{\prime }}^a\phi _{\text{pp},\text{tr}%
}^{a,\sigma \sigma ^{\prime }}({\bf k},\omega )  \label{hipair1} \\
\ \phi _{\text{pp},\text{tr}}^{a,\sigma \sigma ^{\prime }}({\bf k},i\omega
_n) &=&\sum_{{\bf k}^{\prime }{\bf ,}\sigma ^{\prime \prime }\sigma ^{\prime
\prime \prime }}\int_0^\beta d\tau \int_0^\beta d\tau ^{\prime }e^{i\omega
_n(\tau -\tau ^{\prime })}  \nonumber \\
&&\times \langle T[c_{{\bf k,}\sigma }^{\dagger }(\tau )c_{-{\bf k,}\sigma
^{\prime }}^{\dagger }(\tau ^{\prime })c_{{\bf k}^{\prime }{\bf ,}\sigma
^{\prime \prime }}(0)  \nonumber \\
&&\times c_{-{\bf k}^{\prime }{\bf ,}\sigma ^{\prime \prime \prime
}}(0)]\rangle 
\end{eqnarray}
Similar to magnetic susceptibility, $\phi _{\text{pp},\text{tr}}^{a,\sigma
\sigma ^{\prime }}$ depends on $\varepsilon _{{\bf k}}$ and $\omega $ only
and can be generally written as 
\begin{eqnarray}
\phi _{\text{pp},\text{tr}}^{a,\sigma \sigma ^{\prime }}({\bf k},i\omega _n)
&=&A_{\sigma \sigma ^{\prime }}^a\phi _{\text{pp,tr}}(\varepsilon _{{\bf k}%
},i\omega )  \label{fff} \\
\phi _{\text{pp,tr}}(\varepsilon _{{\bf k}},i\omega ) &=&G(i\omega
-\varepsilon _{{\bf k}})G(-i\omega _n-\varepsilon _{{\bf k}})\gamma _{\text{%
tr}}(\varepsilon _{{\bf k}},i\omega )  \nonumber
\end{eqnarray}
where $\gamma _{tr}(\varepsilon _{{\bf k}},i\omega )$ is the pairing vertex.
At $\xi \rightarrow \infty $ we obtain 
\begin{equation}
\phi _{\text{pp},\text{tr}}(\varepsilon _{{\bf k}},i\omega )=F(\varepsilon _{%
{\bf k}},\omega )+F(-\varepsilon _{{\bf k}},\omega )
\end{equation}
where 
\begin{eqnarray}
F(\varepsilon _{{\bf k}},\omega ) &=&\frac{\omega +(M-1)\,\varepsilon _{{\bf %
k}}}{2\,M\,\omega \,\overline{\omega }\varepsilon _{{\bf k}}}\,\left[ 1-%
\frac M2\left( -\frac{M\,\overline{\omega }^2}{2\Delta ^2}\right)
^{M/2}\right.   \nonumber \\
&&\left. \times e^{-\frac{M\,\overline{\omega }^2}{2\,\Delta ^2}}\Gamma
\left( -\frac M2,-\frac{M\overline{\omega }^2}{2\,\Delta ^2}\right) \right]
_{\overline{\omega }=\omega -\varepsilon _{{\bf k}}}\,  \label{FF}
\end{eqnarray}
Note that $\gamma _{tr}$ depends on $\omega $ and $\varepsilon _{{\bf k}}$
separately and $\gamma _{\text{tr}}(0,\omega )=\gamma (\omega ).$ The
pairing susceptibility (\ref{hipair}) is expressed through $\phi _{\text{pp},%
\text{tr}}(\varepsilon _{{\bf k}},i\omega )$ by the relation 
\begin{eqnarray}
\phi _{\text{pp},\text{tr}} &=&\int d\varepsilon \rho _0(\varepsilon )\Phi _{%
\text{tr}}(\varepsilon )  \label{ftr} \\
\ \ \Phi _{\text{tr}}(\varepsilon ) &=&T\sum_{i\omega _n}\phi _{\text{pp},%
\text{tr}}(\varepsilon ,i\omega _n),  \nonumber
\end{eqnarray}
The function $\phi _{\text{pp},\text{tr}}(\varepsilon ,i\omega ),$ which
characterises the contribution of different momenta and energies in the
pairing susceptibility is plotted at $\xi \rightarrow \infty $ in Fig. 4a.
This function is divergent at $\omega \rightarrow 0^{+},$ which signals the
possibility of the triplet pairing at $T\rightarrow 0.$ At finite small $%
\omega $ the function $\phi _{\text{pp},\text{tr}}(\varepsilon ,i\omega )$
is maximum at $\varepsilon =\pm \Delta _0.$ Therefore, contrary to standard
BCS problem, the pairing due to the ground-state FM instability in 2D system
involves particles with finite energy (with respect to the paramagnetic
Fermi surface) and the momenta at the new preformed Fermi surfaces.

At finite $\xi $ the function $\phi _{\text{pp},\text{tr}}(\varepsilon _{%
{\bf k}},i\omega )$ is determined by the Eq. (\ref{fff}); the pairing vertex 
$\gamma _{\text{tr}}$ is obtained from the recursion relation which is
similar to the recursion relation for $\gamma ,$%
\begin{eqnarray}
\gamma _{\text{tr},j}(\varepsilon _{{\bf k}},\omega ) &=&1+\frac{\Delta
_0^2r_j}{2\ln \xi }\int_{-\infty }^\infty \frac{da}{\sqrt{a^2+\xi ^{-2}}} 
\nonumber \\
&&\times \gamma _{\text{tr},j+1}(\varepsilon _{{\bf k}}-v_Fa,\omega
)G_{j+1}(\omega -\varepsilon _{{\bf k}}-v_Fa)  \nonumber \\
&&\times G_{j+1}(-\omega -\varepsilon _{{\bf k}}-v_Fa)  \label{gtrj1}
\end{eqnarray}
with $\gamma _{\text{tr},\infty }(\omega ,\varepsilon _{{\bf k}})=1.$

\begin{figure}[t]
\psfig{file=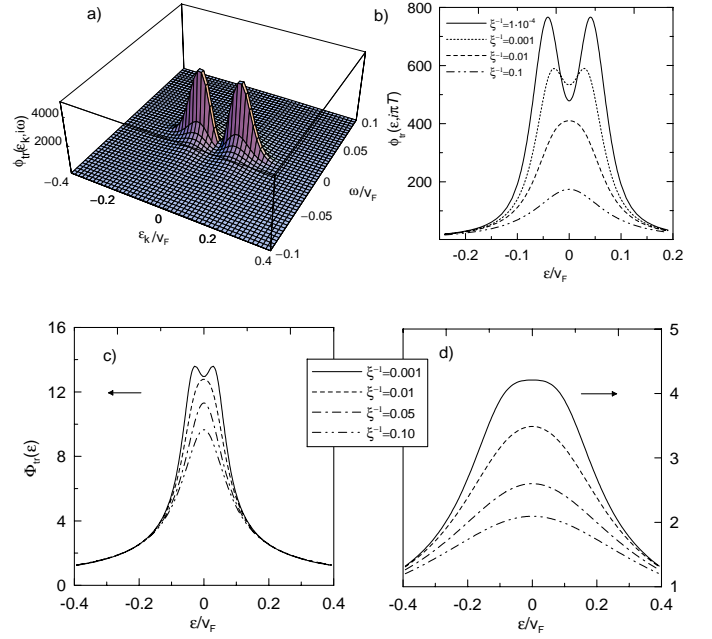,width=90mm,silent=} \vspace{2mm}
\caption{(a) The plot of the function $\phi _{\text{pp},\text{tr}%
}(\varepsilon ,i\omega )$ at $\xi \rightarrow \infty $ and $\Delta _0=0.1v_F$
(b) The function $\phi _{\text{pp},\text{tr}}(\varepsilon ,i\pi T)$ at
finite $\xi $, $\Delta _0=0.1/5^{1/2}v_F$, and $T^{*}=0.05v_F$. (c,d) The
function $\Phi _{\text{tr}}(\varepsilon )$, which determines the triplet
pairing susceptibility according to Eq. (\ref{ftr}), at the same values of $%
\Delta _0$ and $T^{*}$ as in b) (c) and at $\Delta _0=0.1v_F$, $T^{*}=0.25v_F
$ (d).}
\label{fig:Fig4}
\end{figure}
The function $\phi _{\text{pp},\text{tr}}(\varepsilon ,i\pi T)$ at different
values of $\xi $ is shown in Fig. 4b. With decreasing $\xi $ the two-peak
structure of $\phi _{\text{pp},\text{tr}}(\varepsilon ,i\pi T)$ continuously
changes to a one-peak structure at $T=T^{*}/\ln \xi \sim \Delta _0$. The
function $\Phi _{\text{tr}}(\varepsilon )$ for two choices of $T^{*}$ and $%
\Delta _0$ and different values of the correlation length is shown in Figs.
4c,d [we rescale the value of $\Delta _0\propto (T^{*})^{1/2},$ as it
follows from Eq. (\ref{dd})]. Similar to $\phi _{\text{pp},\text{tr}%
}(\varepsilon ,i\pi T),$ the function $\Phi _{\text{tr}}(\varepsilon )$
changes its behavior from the two-peak to a one-peak structure at $T\sim
\Delta _0$, so that the triplet pairing fluctuations are dominating at not
too large $\xi $ at the {\it paramagnetic} Fermi surface.

To clarify the role of the vertex corrections for the triplet pairing, we
plot in Fig. 5 the triplet pairing vertex $\gamma _{\text{tr}}(\varepsilon
,i\pi T)$ at first Matsubara frequency for the same choices of $T^{*}$ and $%
\Delta _0$ as in Fig. 4. We find that at $T^{*}\ll v_F$ (weakly FM ground
state) and sufficiently large correlation length $\ln \xi \gg T^{*}/\Delta
_0 $ the triplet pairing vertex $\gamma _{\text{tr}}(\varepsilon ,i\pi T)$
is considerably enhanced. In particular, we emphasize that the divergence of
the pairing susceptibility at $\xi \rightarrow \infty $ arises {\it solely}
from the vertex corrections.

The triplet pairing vertex $\gamma _{\text{tr}}(\varepsilon ,i\pi T)$ in the
quasistatic approach can be furthermore compared with the result $\gamma _{%
\text{tr}}^E(\varepsilon ,i\pi T)$ of the approach which accounts for the
self-energy corrections only (the analogue of the Eliashberg-type approach
of Refs. \cite{CC,Gorkov}). At $\xi \rightarrow \infty $ we obtain in such
an approach (cf. Ref. \cite{KKI}) 
\begin{eqnarray}
\gamma _{\text{tr}}^E(\varepsilon ,i0^{+}) &=&[1-(M-2)\Delta
_0^2|G_E(\varepsilon +i0^{+})|^2/M]^{-1}  \nonumber \\
G_E(\varepsilon +i0^{+}) &=&[\varepsilon -\Sigma _E(\varepsilon
+i0^{+})+i0^{+}]^{-1}  \nonumber \\
\Sigma _E(\varepsilon +i0^{+}) &=&(\varepsilon -\sqrt{\varepsilon -2\Delta _0%
}\sqrt{\varepsilon +2\Delta _0})/2,  \label{gE}
\end{eqnarray}
the corresponding finite-$\xi $ result can be obtained from Eqs. (\ref{gs1}%
), (\ref{gtrj1}) with $c_j=1$ and $r_j=(M-2)/M$. It can be found from Eq. (%
\ref{gE}) that $\gamma _{\text{tr}}^E(\varepsilon ,i0^{+})<M/2$ and,
therefore, remains finite at $T\rightarrow 0,$ $\xi \rightarrow \infty $. At
the same time, the triplet pairing vertex in the quasistatic approach is
divergent in this limit, leading to the divergence of the triplet pairing
susceptibility. This divergence indicates the possibility of the triplet
pairing due to {\it classical} spin fluctuations, which is complementary to
pairing due to quantum spin fluctuations previously studied in Refs. \cite
{CC,Gorkov}. At $\ln \xi \lesssim T^{*}/\Delta _0$ we find $\gamma _{\text{tr%
}}(\varepsilon ,i\pi T)\simeq 1$ and the vertex corrections are not
important. In this case, the Eliashberg-type approach of Refs. \cite
{CC,Gorkov} is justified.

Therefore, the role of the vertex corrections for the triplet pairing
depends on the temperature and the value of the correlation length, at not
too large correlation length the vertex corrections can be neglected, but
they become crucially important at large $\xi .$

\begin{figure}[t!]
\psfig{file=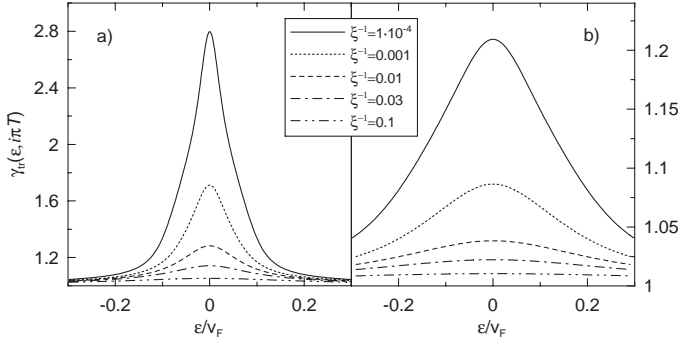,width=90mm,silent=} \vspace{2mm}
\caption{The triplet pairing vertex $\gamma _{\text{tr}}(\varepsilon ,i\pi
T) $ in the quasistatic approach at $\Delta _0=0.1/5^{1/2} v_F$, $T^{*}=0.05
v_F$ (a), $\Delta _0=0.1 v_F$, $T^{*}=0.25 v_F$ (b), and different $\xi $. }
\label{fig:Fig5}
\end{figure}

\section{The quantum-critical regime}

As discussed previously, the static contributions to self-energy and
vertices dominate over quantum contributions for sufficiently large
correlation length, $\xi ^{-1}\ll (T/v_F)^{1/3}.$ Provided that this
inequality is satisfied, one can apply the above consideration to the
quantum-critical regime as well. As was mentioned in Sect. II, in this
regime $\Delta _0\propto T^{1/2}\ln ^{1/2}(v_F/T)$ becomes temperature
dependent itself. The spectral properties in this case depend on the value
of the exponent $\alpha $, which determines the temperature dependence of
the correlation length according to $\xi \sim (T/v_F)^{-\alpha }$. In this
respect, two regimes can be distinguished: (i) $\xi ^{-1}\ll \Delta _0/v_F,$
i.e. $\alpha >1/2$ and (ii) $\Delta _0/v_F\lesssim \xi ^{-1}\ll
(T/v_F)^{1/3},$ i.e. $1/3<\alpha \leq 1/2.$ In the regime (i) spectral
functions have the two-peak structure, so that similar to the RC regime,
studied in Sect. II, the Fermi-surface at finite temperatures is pre-split,
while in the regime (ii) spectral functions have one-peak structure. In both
regimes the real part of the self-energy has positive slope at the Fermi
level $\partial $Re$\Sigma /\partial \varepsilon \sim (T/v_F)\xi ^2$, and
the imaginary part at the Fermi level (associated with the inverse qp
lifetime) is anomalously large, $|$Im$\Sigma |\sim T\xi ,$ so that the qp
picture is invalid. According to Millis theory \cite{Millis}, the
temperature dependence of the correlation length in the QC regime is given
by $\xi ^{-1}\propto T^{1/2}\ln ^{1/2}(v_F/T),$ and therefore this type of
dependence belongs to the regime (ii).

For the charge instability case ($M=1$) the derivative of the real part of
self-energy at the Fermi level is positive, but finite, $\partial $Re$\Sigma
/\partial \varepsilon \sim 1,$ and the imaginary part at the Fermi level $|$%
Im$\Sigma |\sim \Delta _0\sim T^{1/2}\ln ^{1/2}(v_F/T)$ in the regime (i)
and $|$Im$\Sigma |\sim T\xi $ in the regime (ii). Note that $|$Im$\Sigma |$
in the regime (i) does not depend on the value of the exponent $\alpha $ in
this case.

The triplet pairing susceptibility in the QC regime is determined by the Eq.
(\ref{ftr}). The function $\Phi _{tr}(\varepsilon )$ has a one-peak
structure similar to that in the RC regime at not too large correlation
length. According to the results of Sect. III, vertex corrections to the
triplet pairing susceptibility are small at $T\gtrsim \Delta _0$ where the
Eliashberg-type approach of Refs. \cite{CC,Gorkov} is justified. The
corresponding condition in the QC regime coincides, up to logarithmic
corrections, with the condition of the applicability of the susceptibility
ansatz (\ref{hi11}), $T\gtrsim T_X$. At the same time, at $T\ll T_X$ triplet
pairing susceptibility is substantially enhanced over its bare value and
vertex corrections can {\it not} be neglected. The analysis of this case
requires consideration of the nonanalytic corrections to magnetic
susceptibility, which is beyond the scope of present paper. One can expect,
however, that in this regime magnetic and superconducting fluctuations are
strongly coupled and should be considered on the same foot.

\section{Conclusion}

In the present paper we have studied spectral properties and the triplet
pairing fluctuations in the vicinity of a FM instability. The strong FM
fluctuations violate the qp concept near the Fermi level, leading to
anomalously large scattering rate at the Fermi level, $|$Im$\Sigma |\sim
T\xi ,$ and the positive slope of the real part of the self-energy at the
Fermi level, $\partial $Re$\Sigma /\partial \varepsilon \sim (T/v_F)\xi ^2$.
Although these results coincide with the results of the second-order
perturbation theory with respect to coupling of electrons with magnetic
excitations \cite{KKI}, they take into account the self-energy and vertex
corrections. Therefore, these two type of corrections almost compensate each
other, as it was concluded before on the basis of $1/M$ expansion \cite{KKI}%
. At large enough correlation length $\xi ^{-1}\gg \Delta _0/v_F$ ($\Delta _0
$ is the ground state spin splitting in the RC regime and $\Delta _0\propto
T^{1/2}$ in the QC regime) the abovementioned features of the self-energy
lead to the two-peak structure of the spectral function, which implies the
quasisplitting of the Fermi surface, as proposed in Ref. \cite{KKI}. The
structure of the spectral function changes to a one-peak form at $\xi
^{-1}\lesssim \Delta _0/v_F.$ For $M=1$ (charge instability case) the
spectral function has a one-peak structure at arbitrary $\xi $. This does
not restore the qp picture, however, since the slope of Re$\Sigma $ remains
positive, $\partial $Re$\Sigma /\partial \varepsilon \sim 1$ and the
imaginary part $|$Im$\Sigma |\sim \min (\Delta _0,v_F\xi ^{-1})$ is finite
at low $T$.

The triplet pairing susceptibility near the FM instability is considerably
enhanced at low temperatures as compared to its bare value. In the RC regime
at large correlation length the triplet pairing is most strong at the newly
preformed (quasi-split) Fermi-surfaces, and with increasing temperature
(i.e. decreasing correlation length) the triplet pairing arises at the
paramagnetic Fermi surface. The vertex corrections to the triplet pairing
susceptibility are important at low enough temperatures $T\lesssim \max
\{\Delta _0,v_F\xi ^{-1}\}.$

In the quantum critical regime, dynamic contributions with nonzero bosonic
Matsubara frequency to self-energy and vertices can be neglected at low
enough temperatures for $\alpha >1/3,$ where the exponent $\alpha $
describes the temperature dependence of the correlation length, $\xi \sim
(T/v_F)^{-\alpha }.$ Depending on the value of $\alpha ,$ one- or two-peak
structure of the spectral functions is possible, the former arising for $%
1/3<\alpha \leq 1/2,$ the latter for $\alpha >1/2.$ The qp picture is
violated for any $\alpha <1.$ Since, however, the contributions of nonzero
bosonic Matsubara frequencies are different only by power of temperature,
their contribution is expected to be important for a correct quantitative
description of quantum critical regime. The consideration of the triplet
pairing shows that the vertex corrections in the quantum critical regime can
be neglected at not too low temperatures and become non-negligible in the
same temperature range $T\lesssim U^2/v_F,$ where nonanalytic corrections to
magnetic susceptibility become important. The consideration of this region
requires an analysis of magnetic and superconducting fluctuations on the
same foot, which is the subject of future investigations.

In summary, the quasistatic approximation discussed in the present paper
allows for a treatment of the self-energy and vertex corrections which arise
from static magnetic fluctuations. In this respect, such an approximation
has some advantages over $1/M$ expansion, since it does not require $M$ to
be sufficiently large. However, it can be hardly generalized to include
dynamic magnetic contributions with nonzero bosonic Matsubara frequencies.
Therefore, a generalization of the $1/M$ expansion, which includes these
dynamic contributions, is desirable. On the other hand, the generalization
of the quasistatic approach which includes the effect of van Hove
singularities in the electronic spectrum could provide a possibility to
describe qualitatively the properties of real low-dimensional materials.

\section{Appendix. The derivation of the recursion relations at finite
correlation length}

In this Appendix we reconsider the extension of the quasistatic approach to
2D case when the static magnetic susceptibility has the form 
\begin{equation}
\chi ({\bf q},0)=\frac A{q^2+\xi ^{-2}}  \label{hi1}
\end{equation}
The early version of quasistatic approach for 1D models \cite{Sadovskii} can
be directly extended to 2D case only for the factorizable form of the
susceptibility (cf. Refs. \cite{Schmalian,Sadovskii1}) 
\begin{equation}
\chi ({\bf q},0)=A\frac{\xi ^{-1}}{q_{\Vert }^2+\xi ^{-1}}\frac{\xi ^{-1}}{%
q_{\perp }^2+\xi ^{-1}}  \label{hi}
\end{equation}
with $q_{\Vert }$ and $q_{\perp }$ being the components of $q,$ parallel and
perpendicular to the electron momentum ${\bf k}.$ Although the extension of
quasistatic approach to the susceptibility ansatz (\ref{hi1}) was discussed
previously in Ref. \cite{Schmalian}, we argue that this extension does not
treat correctly logarithmic corrections, which arise after integration of
Eq. (\ref{hi1}) over ${\bf q}.$ While these logarithmic corrections are
subleading in the quantum-critical regime, they are crucially important in
the RC regime, where the correlation length is exponentially large.

To discuss the way of a proper generalization of the method, we consider the
contribution of a $2N$-th order diagram for the self-energy (cf. Ref. \cite
{Schmalian}) 
\begin{eqnarray}
\Sigma ^{(2N)}({\bf k},i\omega _n) &=&g^{2N}\sum_{q_1...q_N}\chi ({\bf q}%
_1,0)...\chi ({\bf q}_N,0)  \nonumber \\
&&\times \prod_{j=1}^{2N-1}\left[ i\omega _n-\varepsilon _{{\bf k+}%
\sum_{\alpha =1}^NR_{j\alpha }{\bf q}_\alpha }\right] ^{-1}  \label{See}
\end{eqnarray}
where $g=UT^{1/2}.$ The coefficients $R_{j\alpha }$ determine whether $%
\alpha $-th momentum variable ${\bf q}_\alpha $ enters $j$-th electronic
Green function, see details in Ref. \cite{Schmalian}. At large $\xi \gg 1$
the most important contribution to $\Sigma ^{(2N)}$ comes from small
momenta, and it is sufficient to expand the denominator of Eq. (\ref{See})
in $q.$ For further convenience, we introduce new variables of integration $%
a_\alpha =q_\alpha \cos \theta _\alpha $, where $\alpha =1...N$ ($\theta
_\alpha $ is an angle between ${\bf q}_\alpha $ and ${\bf k}$)$.$ The
integrals over $q$ can be then calculated analytically; using the form of
the susceptibility (\ref{hi1}) we obtain 
\begin{eqnarray}
\Sigma ^{(2N)}({\bf k}_F,i\omega _n) &=&(Ag^2/4\pi )^N\int_{-\infty }^\infty 
\frac{da_1}{\sqrt{a_1^2+\xi ^{-2}}}...\frac{da_N}{\sqrt{a_N^2+\xi ^{-2}}} 
\nonumber \\
&&\times \prod_{j=1}^{2N-1}\left[ i\omega _n-v_F\sum_{\alpha =1}^NR_{j\alpha
}a_\alpha \right] ^{-1}  \label{Sa}
\end{eqnarray}
The corresponding result for the factorized susceptibility ansatz (\ref{hi})
differs by the replacement $\sqrt{a_\alpha ^2+\xi ^{-2}}\rightarrow \xi
(a_\alpha ^2+\xi ^{-2})$ in the denominators of Eq. (\ref{Sa}), $a_\alpha
=q_{\alpha ,\parallel }$ in this case.

For $|\omega |\gg v_F\xi ^{-1}$ one can neglect $a_\alpha $ in the
denominators of Green functions in Eq. (\ref{Sa}) to obtain 
\begin{eqnarray}
\Sigma ^{(2N)}({\bf k}_F,\omega ) &\simeq &(Ag^2/4\pi )^N\omega
^{-(2N-1)}\ln ^N(\xi \omega /v_F),\,\,  \nonumber \\
|\omega | &\gg &v_F\xi ^{-1}  \label{a1}
\end{eqnarray}
To find asympthotic form of the self-energy at small $|\omega |\ll v_F\xi
^{-1}$, we shift contours of integrations in Eq. (\ref{Sa}) to the upper
half of the complex plane. The integrals are then determined by the
contributions of branch cuts of square roots and 
\begin{eqnarray}
\Sigma ^{(2N)}({\bf k}_F,\omega ) &\simeq &i(Ag^2/4\pi )^N\xi
^{2N-1}f(\{R_{j\alpha }\}),  \nonumber \\
|\omega | &\ll &v_F\xi ^{-1}.  \label{a2}
\end{eqnarray}
where $f(\{R_{j\alpha }\})$ is some function which depends on the
coeffitients $R_{j\alpha }$ only. One can see, that at $\,|\omega |\ll
v_F\xi ^{-1}$ the self-energy $\Sigma ^{(2N)}({\bf k}_F,\omega )$ {\it does
not }acquire logarithmic corrections. At the same time, the approach of Ref. 
\cite{Schmalian} leads to logarithmic corrections in the self-energy in both%
{\it \ }the{\it \ }limits, $|\omega |\gg v_F\xi ^{-1}$ {\it and} $|\omega
|\ll v_F\xi ^{-1},$ due to an incorrect factorization of Bessel functions of
sums of auxiliary variables, used in Ref. \cite{Schmalian}. Note, that for
the ansatz (\ref{hi}), branch cut singularities of the integrands in Eq. (%
\ref{Sa}) are replaced by single poles, so that at arbitrary $|\omega |\ll
v_F$ we obtain 
\begin{equation}
\Sigma ^{(2N)}({\bf k},\omega )\simeq (Ag^2/4\pi )^N\prod_{j=1}^{2N-1}\frac 1%
{\omega -\varepsilon _{{\bf k}}+in_jv_F\xi ^{-1}},  \label{aa1}
\end{equation}
with $n_j=\sum_{\alpha =1}^NR_{j\alpha }$, which reproduces the result of
Refs. \cite{Schmalian,Sadovskii1}.

For the form of susceptibility (\ref{hi1}) one can develop an approximate
approach, which becomes exact at $|\omega |\gg v_F\xi ^{-1}.$ Similar to
Refs. \cite{Sadovskii,Sadovskii1} we approximate the contribution of any
diagram by the contribution of corresponding noncrossing diagram. Although
the multiplicity factors are the same, as derived in Ref. \cite{Schmalian},
the expression for the corresponding noncrossing diagram is {\it different}.
Indeed, substituting the dressed Green function instead of the bare one in
Eq. (\ref{Sa}) with $N=1$, and taking into account that the self-energy
depends on $\omega -\varepsilon _{{\bf k}}$ only, we obtain the recursion
relation

\begin{eqnarray}
\Sigma _j(\omega ) &=&\frac{Ag^2c_j}{4\pi }\int_{-\infty }^\infty \frac{da}{%
\sqrt{a^2+\xi ^{-2}}}  \label{ss} \\
&&\times \frac 1{\omega -v_Fa-\Sigma _{j+1}(\omega -v_Fa)}  \nonumber
\end{eqnarray}
where $c_j=j/M$ (even $j$) and $c_j=(j+M-1)/M$ (odd $j$), Contrary to Ref. 
\cite{Schmalian}, this is an integral rather than an algebraic relation. The
initial condition for Eq. (\ref{ss}) is $\Sigma _\infty (\omega )=0,$ the
self-energy is given by $\Sigma (\omega )=\Sigma _1(\omega ).$ For the
vertices we obtain similarly

\begin{eqnarray}
\gamma _j(\omega ) &=&1-\frac{Ag^2r_j}{4\pi }\int_{-\infty }^\infty \frac{da%
}{\sqrt{a^2+\xi ^{-2}}}  \nonumber \\
&&\times \frac{\gamma _{j+1}(\omega -v_Fa)}{[\omega -v_Fa-\Sigma
_{j+1}(\omega -v_Fa)]^2}  \label{gj} \\
\gamma _{\text{tr},j}(\varepsilon _{{\bf k}},\omega ) &=&1+\frac{Ag^2r_j}{%
4\pi }\int_{-\infty }^\infty \frac{da}{\sqrt{a^2+\xi ^{-2}}}  \nonumber \\
&&\times \gamma _{\text{tr},j+1}(\varepsilon _{{\bf k}}-v_Fa,\omega )[\omega
-\varepsilon _{{\bf k}}-v_Fa  \nonumber \\
&&-\Sigma _{j+1}(\omega -\varepsilon _{{\bf k}}-v_Fa)]^{-1}[-\omega
-\varepsilon _{{\bf k}}  \nonumber \\
&&-v_Fa-\Sigma _{j+1}(-\omega -\varepsilon _{{\bf k}}-v_Fa)]^{-1}
\label{gtrj}
\end{eqnarray}
with $r_j=j/(M-2)$ (even $j$), $r_j=(M-2)(j+M-1)/M^2$ (odd $j$) and $\gamma
_\infty (\omega )=\gamma _{\text{tr,}\infty }(\omega )=1.$

As mentioned above, for the ansatz (\ref{hi}) the replacement $\sqrt{%
a_\alpha ^2+\xi ^{-2}}\rightarrow a_\alpha ^2\xi +\xi ^{-1}$ in Eqs. (\ref
{ss})-(\ref{gtrj}) should be made. The integrals in the Eqs. (\ref{ss}) and (%
\ref{gj}) can be then evaluated analytically, leading to the recursion
relations of Refs. \cite{Schmalian,Sadovskii}. At the same time, the
integrating expression of Eq. (\ref{gtrj}) is nonanalytical in both, upper
and lower half-plane, and therefore can not be reduced to an algebraic form
even for the factorizable susceptibility ansatz (\ref{hi}).

\section{Acknowledgements}

I am grateful to A. P. Kampf for helpful discussions and W. Metzner and V.
Yu. Irkhin for many valuable comments. This work was partially supported by
Grant No.~747.2003.2 (Support of Scientific Schools) from the Russian Basic
Research Foundation.


\begin{references}
\bibitem{ARPES}  A. Damascelli, Z. Hussain, and Z.-X. Shen, Rev. Mod. Phys. 
{\bf 75}, 473 (2003).

\bibitem{Pines}  P. Monthoux, A. V. Balatsky, and D. Pines, Phys. Rev. Lett. 
{\bf 67}, 3448 (1991); Phys. Rev. B {\bf 46}, 14803 (1992).

\bibitem{Schmalian}  J. Schmalian, D. Pines, and B. Stojkovic, Phys. Rev.
Lett. {\bf 80}, 3839 (1998); Phys. Rev. B {\bf 60}, 667 (1999).

\bibitem{SF}  Ar. Abanov, A. V. Chubukov, and J. Schmalian, Adv. Phys. {\bf %
52}, 119 (2003);

\bibitem{FLEX}  J. J. Deisz, D. W. Hess, and J. W. Serene, Phys. Rev. Lett. 
{\bf 76}, 1312 (1996); J. Altmann, W. Brenig, and A.P. Kampf, Eur. Phys. J.
B {\bf 18}, 429 (2000).

\bibitem{TPSC}  J. Vilk and A.-M. S. Tremblay, J. Phys. I {\bf 7}, 1309
(1997); B. Kyung, Phys. Rev. B {\bf 58}, 16032 (1998).

\bibitem{DCA}  C. Huscroft, M. Jarrell, Th. Maier, S. Moukouri, and A. N.
Tahvildarzadeh, Phys. Rev. Lett. {\bf 86}, 139 (2001).

\bibitem{KK}  A. A. Katanin and A. P. Kampf, Phys. Rev. Lett. {\bf 93},
106406 (2004); D. Rohe and W. Metzner, Phys. Rev. B {\bf 71}, 115116 (2005).

\bibitem{Lee}  N. Nagaosa and P. A. Lee, Phys. Rev. Lett. {\bf 64}, 2450
(1990); P. A. Lee and N. Nagaosa, Phys. Rev. B {\bf 46}, 5621 (1992) .

\bibitem{Castellani}  C. Castellani, C. Di Castro, and M. Grilli, Phys. Rev.
Lett. {\bf 75}, 4650 (1995); C. Castellani, S. Caprara, C. Di Castro, and A.
Maccarone, Nucl. Phys. B {\bf 594}, 747 (2001).

\bibitem{MetznerSE}  W. Metzner, D. Rohe, and S. Andergassen, Phys. Rev.
Lett. {\bf 91}, 066402 (2003).

\bibitem{Sachdev}  S. Sachdev, {\it Quantum phase transitions}, Cambridge
University Press, 1999.

\bibitem{TPSC1}  V. Hankevych, B. Kyung, and A.-M.S. Tremblay, Phys. Rev. B 
{\bf 68}, 214405 (2003).

\bibitem{KKI}  A. A. Katanin, A. P. Kampf, and V. Yu. Irkhin, Phys. Rev. B 
{\bf 71}, 085105 (2005).

\bibitem{Bedell}  K. B. Blagoev, J. R. Engelbrecht, and K. S. Bedell, Phys.
Rev. Lett. {\bf 82}, 133 (1999); Z. Wang, W. Mao, and K. Bedell, Phys. Rev.
Lett. {\bf 87}, 257001 (2001).

\bibitem{CC}  A. V. Chubukov, A. M. Finkel'stein, R. Haslinger, and D. K.
Morr, Phys. Rev. Lett. 90, 077002 (2003).

\bibitem{Gorkov}  M. Dzero and L. P. Gor'kov, Phys. Rev. B {\bf 69}, 092501
(2004).

\bibitem{Sadovskii}  M. V. Sadovskii, Zh. Eksp. Theor. Phys. {\bf 77}, 2070
(1979) [Sov. Phys. JETP {\bf 50}, 989 (1979)]

\bibitem{Sadovskii1}  E. Z. Kuchinskii and M. V. Sadovskii, Zh. Eksp. Theor.
Phys. {\bf 115}, 1765 (1999) [Sov. Phys. JETP {\bf 88}, 968 (1999)]; M. V.
Sadovskii, Usp. Fiz. Nauk. {\bf 171}, 539 (2001) [Physics-Uspekhi {\bf 44},
515 (2001)]; cond-mat/0408489 (unpublished).

\bibitem{Month}  P. Monthoux, Phys. Rev. B {\bf 68}, 064408 (2003).

\bibitem{Nonan}  D. Belitz, T. R. Kirkpatrick, and T. Vojta, Phys. Rev. B 
{\bf 55}, 9452 (1997); A. V. Chubukov and D. L. Maslov, Phys. Rev. B {\bf 68}%
, 155113 (2003); A. V. Chubukov, C. Pepin, and J. Rech, Phys. Rev. Lett. 
{\bf 92}, 147003 (2004).

\bibitem{OurP}  P. Igoshev, A. A. Katanin, and V. Yu. Irkhin, to be
published.

\bibitem{HertzEdw}  J. A. Hertz and D. M. Edwards, J. Phys. F {\bf 3}, 2174
(1973), ibid. F {\bf 3}, 2191 (1973).

\bibitem{CSY}  A. V. Chubukov, S. Sachdev, and J. Ye, Phys. Rev. B {\bf 49},
11919 (1994).

\bibitem{Quasi2D}  V. Yu. Irkhin and A. A. Katanin, Phys. Rev. B {\bf 55},
12318 (1997).

\bibitem{Schutz}  F. Schuetz, L. Bartosch, and P. Kopietz, cond-mat/0409404
(unpublished).

\bibitem{MyWard}  A. A. Katanin, Phys. Rev. B {\bf 70}, 115109 (2004).

\bibitem{Millis}  A. J. Millis, Phys. Rev. B{\bf 48}, 7183 (1993).
\end{references}
\end{document}